# QoS Based Framework for Effective Web Services in Cloud Computing

**DebajyotiMukhopadhyay, Falguni J. Chathly, Nagesh N. Jadhav**

Department of Information Technology, Maharashtra Institute of Technology, Pune 411038, India
Email: {debajyoti.mukhopadhyay, chathly.falguni, nagesh10}@ gmail.com

**ABSTRACT**

Enhancements in technology always follow Consumer requirements. Consumer requires best of service with least possible mismatch and on time. Numerous applications available today are based on Web Services and Cloud Computing. Recently, there exist many Web Services with similar functional characteristics. Choosing "a-right" Service from group of similar Web Service is a complicated task for Service Consumer. In that case, Service Consumer can discover the required Web Service using non functional attributes of the Web Services such as QoS. Proposed layered architecture and Web Service-Cloud i.e.WS-Cloud computing Framework synthesizes the Non functional attributes that includes reliability, availability, response time, latency etc. The Service Consumer is projected to provide the QoS requirements as part of Service discovery query. This framework will discover and filter the Web Services form the cloud and rank them according to Service Consumer preferences to facilitate Service on time.

## 1. Introduction

Following the global economic recession, businesses are gradually looking for more innovative ways to cut technical costs while maximizing value, to acquire competitive hold on IT market. Growing acceptance of pioneering technologies makes cloud computing, the biggest buzzword in IT. Economically, cloud computing appeals that customers only use what they need, and only pay for what they actually use.

Cloud Computing is a upcoming model of convenient, on-demand communication and symbol of Internet, representing complex infrastructure, including configurable computing resources such as, software, hardware, processing power, applications and storage, as a Service, among many computers. A Cloud is a large group of interconnected computers that extends beyond a single company or enterprise [1-2].Such computing helps cloud customers to expand their local computing power onto the relatively infinite processing power of the Internet. Simple cloud is shown in Figure 1.

Web Services can be defined as a technology for offering software Services or general-purpose architecture that will trigger an essential shift in the way that all distributed systems are created. Services provide strong interface for collection of operations being accessed on network. This architecture works with two main entities, the one who launches particular Service, the Service Provider and the one who consumes Service, the Service Consumer. Web Services share business logic, data and processes through a programmatic interface across a network. Developers can then add the Web Service to a GUI to offer specific functionality to users. Web Services allow different applications from different sources to communicate with each other without time-consuming custom coding, and because all communication is in XML, Web Services are not tied to any operating system or programming language.

There are many ways that a requester entity might engage and use a Web Service. In general, the Service requester and Service Provider entities become known to each other (or at least one becomes know to the other); then requester and Provider entities somehow agree on the Service description and semantics that will govern the interaction between the requester and Provider agents; once both agrees Service description and semantics are realized by the requester and Provideragents;and the requester and Provider agents exchange messages, thus performing some task on behalf of the requester and Provider entities. (i.e., the exchange of messages with the Provider agent represents the concrete manifestation of interacting with the Provider entity's Web Service.)

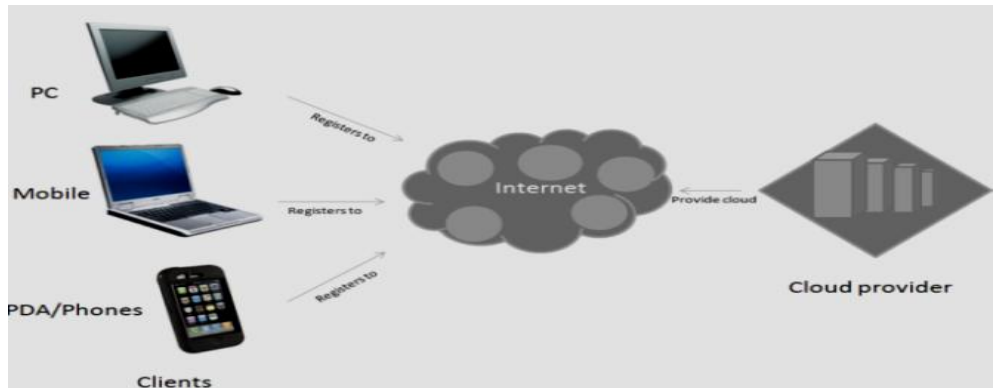
**Figure 1 Simple Cloud**

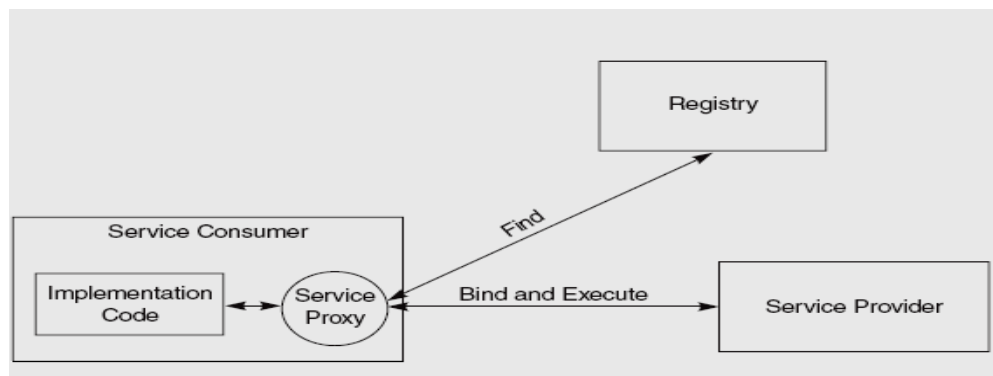
**Figure 2 Service Binding**

The process of agreement between Requester and Provider is Service binding. Refer Figure2.

## 2. Background Research
### 2.1. Existing Systems
To remove vulnerabilities with single processor, single database and single node architecture, the cloud computing research is getting velocity to the fullest. Not only industries, also academia, are actively participating in finding appropriate solutions. In [3], IBM raised an issue, with single user cloud Provider, that has limited resource to use, and has lack of interoperability among cloud Providers also, prevents deployment across different clouds. A.NET based cloud computing a software platform named Aneka [4], and Reservoir architecture; the computational resources within a site are partitioned by a virtualization layer into virtual execution environments (VEEs) is used for clouds. In [5], Huang and team from IBM described a service oriented cloud computing platform that enables web-delivery of application-based services with a set of common business and operational services. Abicloud, Eucalyptus, Nimbu and OpenNebula are among those available clouds. Similarly, different types of clouds and their varying interoperability have been developed.

The proposed architecture concentrate on agent based computing including QoS parameters. The key component of this architecture is Service Discovery Module, which is consisted of different internal working blocks to help in verdict the best of available service, as in application layer that contains VM (Virtual Machines) monitor, Service Request Monitor, and Request Dispatcher. The Dispatcher routes the requests from users/brokers to the cloud resources which is based on the feedback from VM and Service Request monitors that can fulfill their QoS requirements.

### 2.2. Issues with Current Clouds
Although, following best practices in accordance to requirement and trends in need, current cloud computing architectures, have flaws with effective working and service on demand parameters. Some are mentioned as follows:

- Lack of Technicality

When cloud is built with less technical knowledge, more complex computing rules and without technical layout, it leads to dangerous outcomes. Thus, proper designing and effective productive layout is very important [6].Service Components on Cloud Computing are tightly coupled. This is the initial problem with any Cloud Client. This can be noticeably explained using an analogy. A person that buys laptop can rather buy from well renowned company or start to assemble it. The computing resources available over the internet, current cloud implementations do not allow kind of flexibility where user can enjoy facilities of all Providers. If a customer opts to use Amazon S3 storage service, he is then stuck with other cloud computing services Amazon provides, such as EC2, Elastic Map Reduce [7]. Even though sincere cost for a cloud computing deployment gets compact and long term lease is eliminated, more attempt and money is spent on setting up the application for a precise cloud platform which makes it tricky to migrate the same application onto a different cloud. Often, application migration simply may mean redevelopment. For example, applications deployed on Amazon EC2 cannot be migrated easily due its particular storage framework [8].

- Need of Effective Initial Investment

If the initial investment of every project is undersized, it is probable that loads of projects can be launched in haste. Finally, its result is variable, so the requirement cannot match it. The management cost rises sharply. It is noticeably saw that the cloud computing do solve existing questions, also rises new challenges to meet. Cloud Computing balances old problems and new requirements.

- Privacy Maintenance

The most perturbing question is the privacy on cloud computing. The most all the rage business application based on internet is insightful information like salary and client account management. This happens often, and every time when there is data loss [9].

## 2.3. Literature Review

The Agent of Cloud is intermediate entity that binds the values in terms of Services provided and needed. [10-12] Cloud Agent does the job of searching Web Service from the Service database on Cloud and the components are described further refer figure 3.

- Web Interface

Web interface is the graphical user interface provided for registration of Services to cloud Service database and querying the cloud Service database to discover required Service. Input query includes type of Service, keywords, technical specifications, cost and time requirements.

- Service Discovery Agent

Service discovery agent basically consist of query processing module, Service reasoning module, price and time slot matching module, and Service rating module.

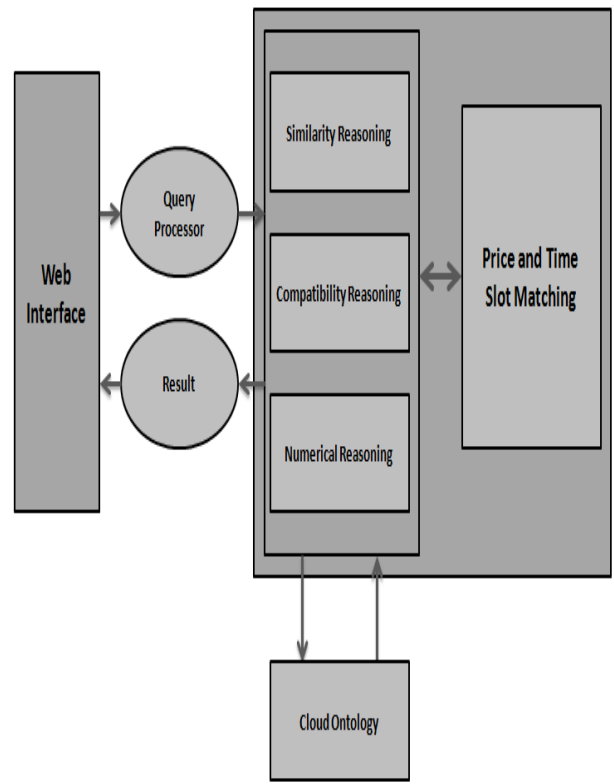

**Figure 3 Cloud Agent Architecture**

- Cloud Service Database

Cloud Service Database holds all the registered Services.

- Query Processor

Once user enters the Service preferences in the web interface query processor processes the query on cloud Service database and return set of Services.

- Cloud Ontology

Cloud Ontology is used for similarity, compatibility and numerical reasoning. Result returned by the query processor is then filtered and ranked accordingly. OWL-S editor can be used to generate the Ontology. Filtered Services are then matched against the price and time slot information provided by Service Consumer.

## 3. Proposed LayeredArchitecture

The proposed 4-tier WS-Cloud computing layered architecture adds spontaneity to the existing architecture. Figure 4 shows architecture design of proposed four-tier architecture.

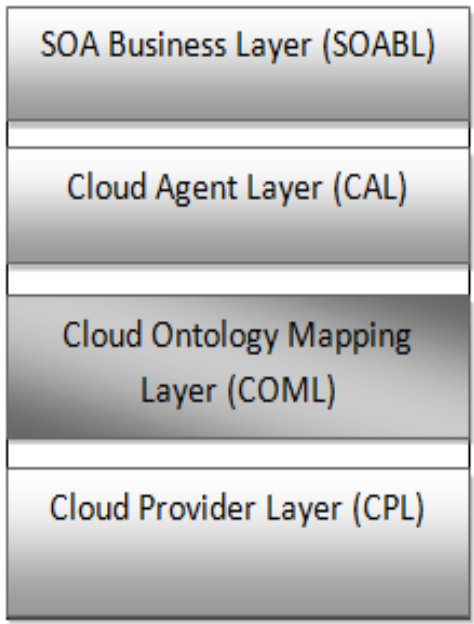

**Figure 4 Layered Architecture**

Business specifications and IT market requirement needs better and efficient applicability of cloud. Cloud is offered in terms of different layers. Layers are the intermediate increments towards successful working. Agent acts as converters or processors of in between queries and proposes new way of working using mapping of storage spaces and computing powers.

- Cloud Provider Layer (CPL)

This layer consists of collection of clouds and it is responsible for providing cloud Services. Individual Cloud Provider Layer: Current cloud implementations are shown in this layer. Cloud Providers organizes their own data centers that rules the Services offered by Providers. Technology of proprietary virtualization is used where each cloud may have its technology setup or it employ open source virtualization technology, such as Eucalyptus [13]. Also, Market-Oriented Cloud Architecture can be used [14], within each individual cloud. Internally, when interoperability is considered, there is a request from Dispatcher working with Virtual Machine Monitor and Service/App Governance Service to allocate the requests to the available recourses. This layer includes components like Storage Service, for common shared storage data, Computing Service, managing effective deliverability to Cloud Consumers and Communication Service, for easy transfer of data throughout, having standardized interfaces, so they can be combined with services from other cloud Providers to build a cross-platform virtual computer on the clouds.

- Cloud Ontology Mapping Layer (COML)

As several clouds exists and each one is responsible for providing specific Services, difficulties in migrating cloud application from one cloud to another cloud is one of the major drawbacks of cloud computing. This can be achieved with the help of cloud Ontology mapping layer. It allows migration of cloud application to another cloud. For this COML uses several Ontology systems. See Figure 5

- Storage Ontology

Storage Ontology is responsible for data manipulation in clouds. It includes select, insert, delete and update operations on data.

- Computing Ontology

Distributed computing on clouds is defined at this level.

- I/O Ontology

It provides Ontology for communication among different clouds.

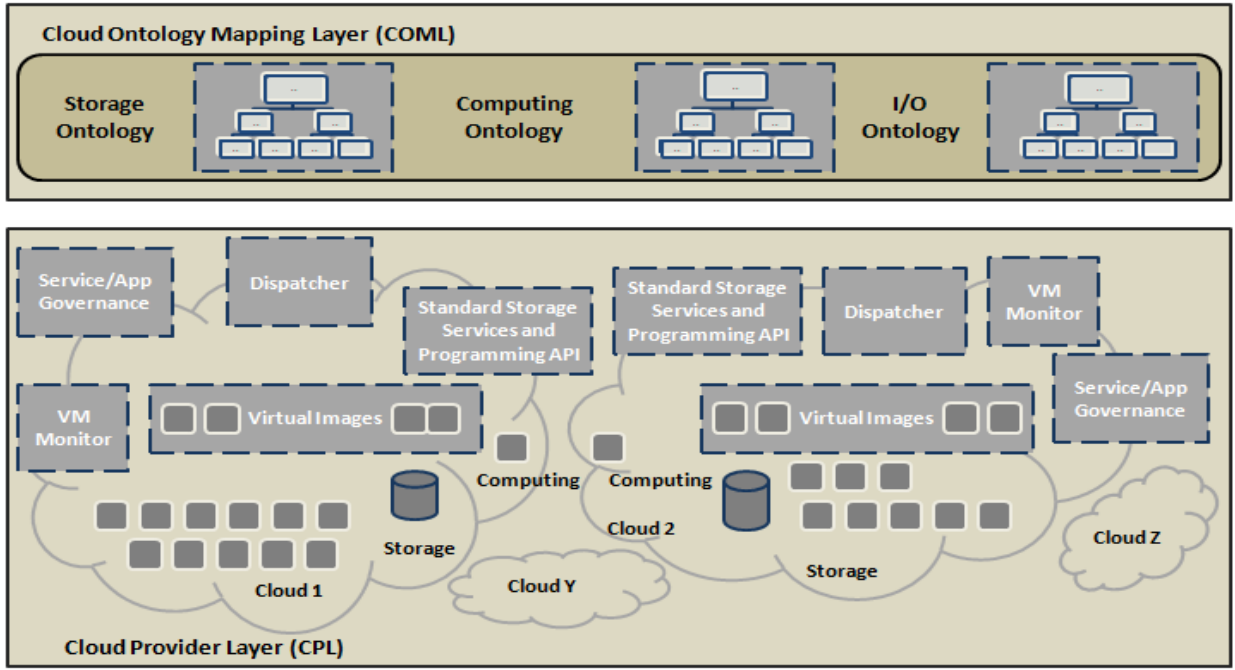

**Figure 5 COML and CPL Architecture**

- Cloud Agent Layer (CAL)

Each cloud Service Provider has agent or broker associated with it. Cloud Agent serves as mediator between Service Oriented Architecture (SOA) and cloud Service Provider. Tasks performed by Cloud Agent are enlisted below [15]. See Figure 6

- Information Publishing

Each cloud Provider publishes Service specifications and its pricing information to the cloud agents. Information includes Service Provider's company name, company address, website and contact information. It also includes resource type and its specifications. Resource type can be computer, storage or communication. More importantly it includes pricing information which may vary from Service Provider to Provider.

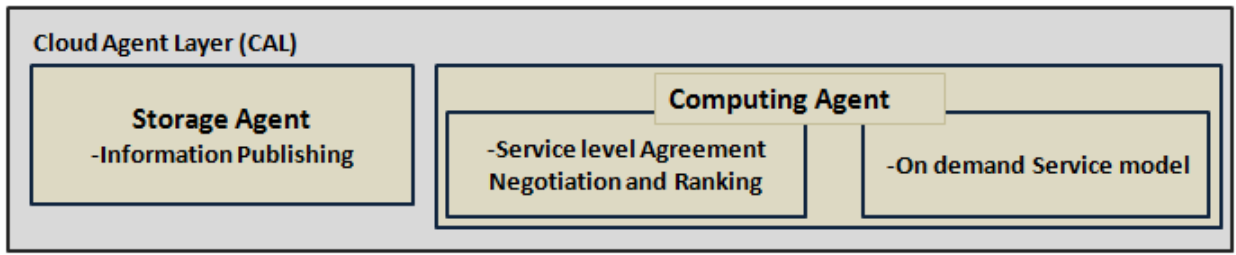

**Fig. 6 CAL Architecture**

- Service level Agreement Negotiation and Ranking

Based on QoS agent ranks published cloud resources. QoS includes reliability, availability, price, latency, security, compliance etc. Agent also provides dynamic Service level agreement between cloud user and cloud Service Provider.

- On demand Service model

As the demand for Services is periodic and seasonal, on demand Service is an important aspect of cloud computing. Exact demand prediction and provision is a critical issue in cloud computing.

- SOA Business Layer (SOABL)

SOA Business layer, is the most interactive and live layer, throughout.

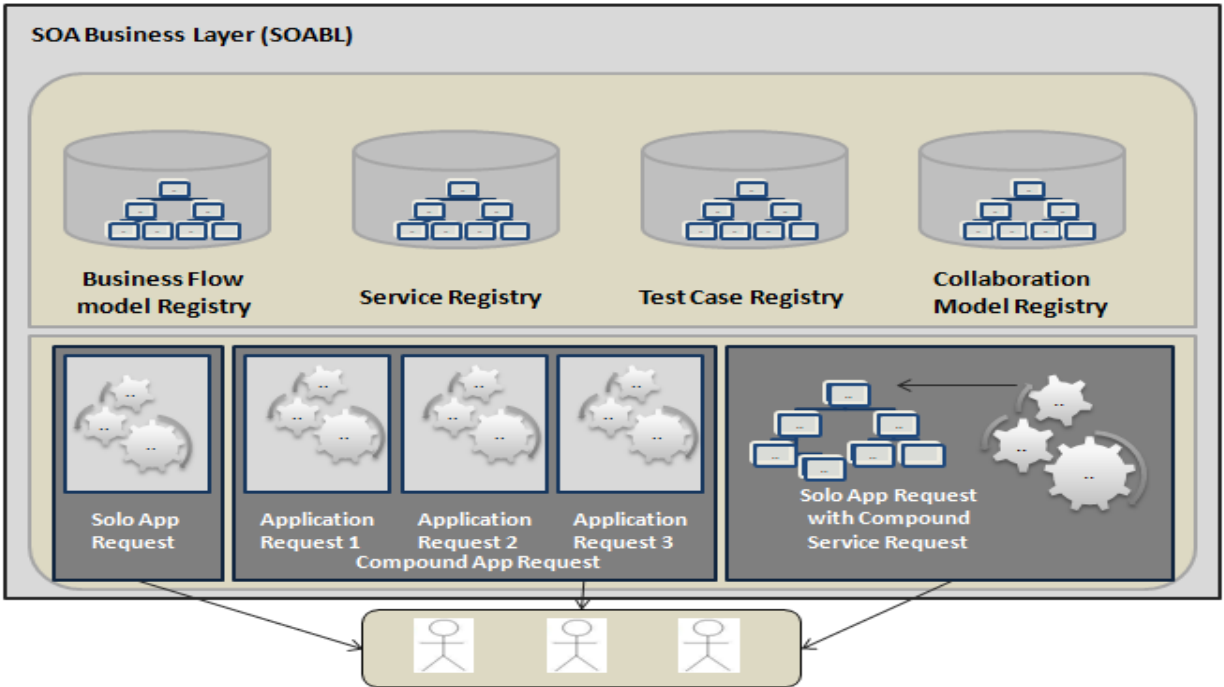

**Figure 7 SOABL Architecture**

All business rules, contracts and agreements are communicated through this layer directly. All the required and available services are stored, collaborated and tested on this stage. It consists of Published Services as well as other artifacts. Different kinds of registries like Business Flow model Registry, in which modeled business rules and commitments are maintained and updated. Service Registry that holds all Published Services or address of Service Providers. Test Case Registry includes the set of service- testing tools, to make Service liable to use and delivered. Collaboration Model Registry isthe most important type, because it decides the actual Service from set of available services from different cloud arrangements.All the information is published into registry which is indexed according to Ontology. In Service Oriented cloud computing architecture Services are published as deployable packages, which can be easily redeployed and replicated on different clouds. Application developers can select the cloud on which they want to run their Services. Registered Service Users, can inquire for Solo Applications or can use Compound i.e. more than one Applications and could even request for both. It is shown in Figure 7.

## 4. Proposed QoS based Framework for WS-Cloud Computing

Newly, there exist many Web Services with similar functional characteristics. Choosing "a-right" Service from group of similar Web Service is a complicated task for Service Consumer. In that case, Service Consumer can discover the required Web Service using non functional attributes of the Web Services such as QoS. Non functional attributes include reliability, availability, response time, latency etc. The Service Consumer is projected to provide the QoS requirements as part of Service discovery query. Our proposed framework will discover and filter the Web Services form the cloud and rank them according to Service Consumer preferences. Framework is shown in Figure 8.

### 4.1 Steps for Service Registration and Discovery on WS-Cloud

In general, there exist two different techniques for accessing web services: using SOAP (Simple Object Access Protocol) and the REST (Representational State Transfer) approach. SOAP provides the mechanism that allows access to objects across the heterogeneous network. SOAP defines the XML-based information which can be used for exchanging ordered and typed information in a decentralized distributed environment between peers, including Web Services and UDDI registries that represent service brokers through which Providers advertise their services [16]. The Service Provider supplies a Service proxy to the Service Consumer. SOAP Request and SOAP Response are the ways through which SOAP files are bound to send request and get response. The route of Web Service invocation initiates, when the client-side proxy hush-up the user's request (method call) into a SOAP message and sends it to the service, which extracts the call from the received message, and then executes the call to produce the relative results, wraps the results into a SOAP message, and sends it to the client. Upon receiving the message, the same proxy extracts the results and hands them over to the calling client application.The Service Consumer executes the request by calling an API function on the proxy. The service proxy, shown in Figure 8, finds a contract and a reference to the Service Provider in the registry. It then formats the request message and executes the request on behalf of the consumer. The service proxy is a convenience entity for the Service Consumer. It is not mandatory; the Service Consumer developer could write the necessary software for accessing the service directly. The service proxy can enhance performance by caching remote references and data. When a proxy caches a remote reference, subsequent service calls will not require additional registry calls. By storing service contracts locally, the consumer reduces the number of network hops required to execute the service. In addition, proxiescan improve performance by eliminating network calls altogetherby performing some functions locally.

For service methods that do not require service data, the entire method can be implemented locally in the proxy. Methods such as currency conversion, tip calculators, and so on, can be implemented entirely in the proxy. If a method requires some small amount of service data, the proxy could download the small amount of data once and use it for subsequent method calls. The fact that the method is executed in the proxy rather than being sent to the service for execution is transparent to the Service Consumer. However, when using this technique it is important that the proxy support only methods the service itself provides. The proxy design pattern states that the proxy is simply a local reference to a remote object. If the proxy in any way changes the interface of the remote service, then technically, it is no longer a proxy. A Service Provider will provide proxies for many different environments. A service proxy is written in the native language of the Service Consumer. For instance, a Service Provider may distribute proxies for Java, Visual Basic, and Delphi if those are the most likely platforms for Service Consumers. Although the service proxy is not required, it can greatly improve both convenience and performance for Service Consumers. Before an application can begin communicating with a service, it must first discover it and get its specifics (supported methods and invocation details), and then generate its proxy. The proxy allows a client application to make method calls as though it were calling a local function. The proxy is fashioned by first generating a source file from the Service's Web Service Description Language (WSDL) file, which describes the Web Service, how to access it, the operations it performs, the types of parameters to be passed to each of the supported methods, and the types of returned results. After the source file is generated, it is compiled into a proxy class that is finally registered with the client application [16].

**1**. Registration Process

Web Service Provider registers its Services or facilities in to Service Discovery Module. Proposed module includes the process of publishing Services into UDDI Registry through Service Publisher Block (SPB). SPB is responsible for checksum of authenticity of QoS parameters entered by Service Provider. SPB validates the QoS parameters and generate the certificate for each of the registered Service. All the validated Services are then published with UDDI registry.A Cloud computing system consists of a collection of inter-connected and virtualized computers dynamically provisioned as one or more unified computing resource(s) through negotiation of service-level agreements (SLAs) between Providers and consumers. Certificate generated by SPB is sent back to Service Provider for their later reference. QoS parameters related to each registered Service gets stored into tModel in UDDIregistry.

2. Discovery process

Discovery process is meant to find appropriate Service from log of Services available. The Service Consumer can search the UDDI registry for a specific Service through the Service Discovery Module. The Service Consumer requests Web Service through Web Interface specifying QoS requirements in the Query. Query gets internally converted into XML.

Query is then passed to Service Name Matching Block (SNMB) which extracts the Web Services based on their name from the UDDI registry. SNMB is responsible for similarity, compatibility and numerical reasoning. Only after passing all the stages of the SNMB, the refined Services are passed to Service Consumer. The feedback stream is maintained for future confirmation. It uses OWL for selected Service reasoning. Matched list of Web Services are then given to QoS Parameter Matching

Block (QPMB), whose job is to rank and filter the matched Services based on the QoS Services preferences given by Service Consumer. Refer Table 1 and Table 2.

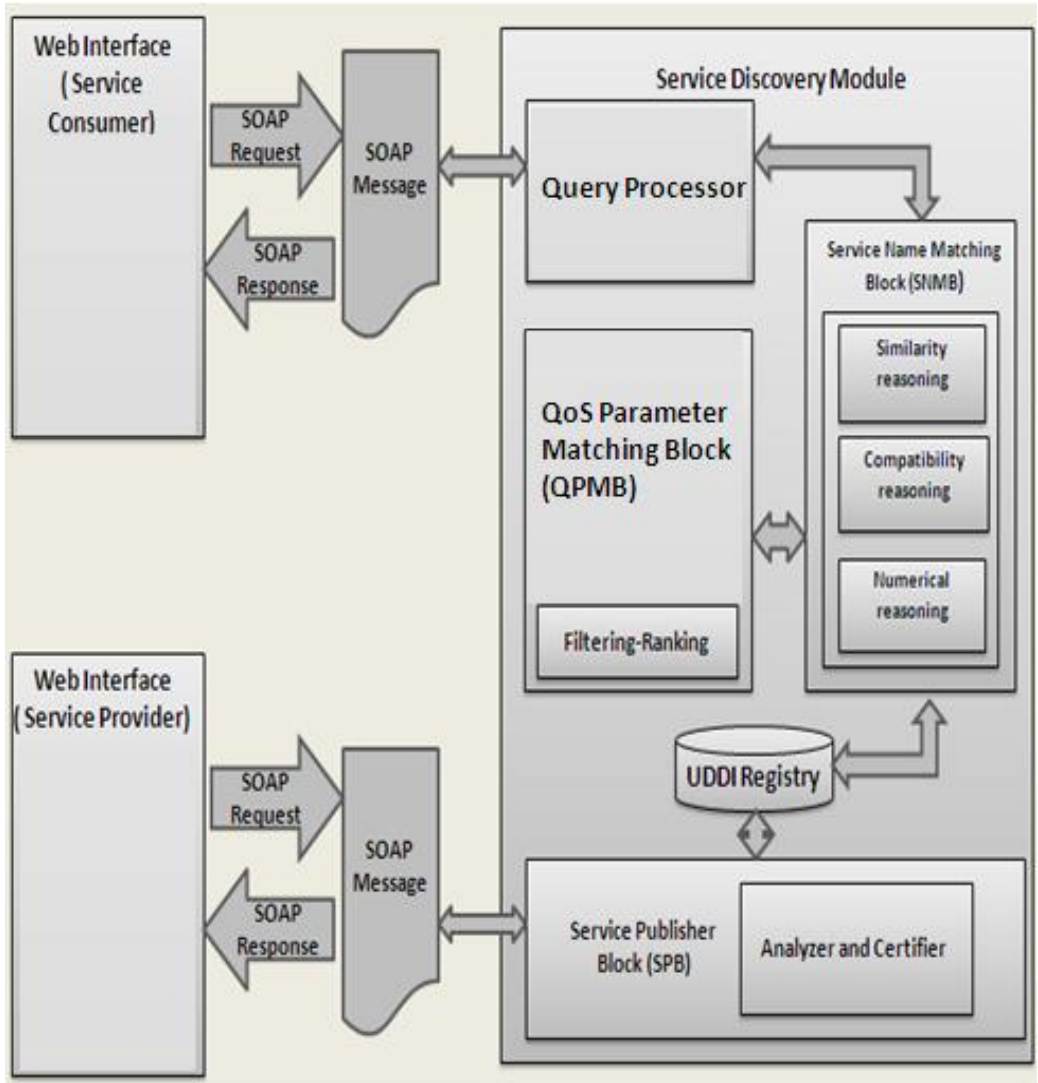

**Figure 8 QoS based Framework for WS-Cloud Computing**

3. Sample Pseudo Code

Table 1 includes the pseudo code for Web Service discovery using QoS parameters. Table 2 provides description for steps to be carried out while performing Web Service discovery process

**Table 1 Web Service Discovery Pseudo Code**

| Discovery Process |
|---|
| **Input** Service to be discovered |
| discoverService( QoS specifications, QoSpref, Service name) |
| **Output** Appropriate matched Service |
| discoverService( QoS specifications, QoSpref, Service name) |
| { Matches=searchUDDI(Service name); |
|   if Matches FOUND |
|     {   QMatch=SearchQoS(Matches, QoS specifications); |

```
    {   If QMatch Found
            {
finalmatch=RankServices(QMatch,, QoSpref);
                return finalmatch;
            }
        }
    }
```

**Table 2 Steps for WS-Cloud discovery on cloud**

| Step | |
|---|---|
| Step 1. | Service Consumer forwards request using Web Interface on to cloud. |
| Step 2. | Request reaches to Service Discovery Module using SOAP messages. |
| Step 3. | Query processor processes the source file into XML format and forwards to SNMB. |
| Step 4. | Internally, Query gets validated through QPMB where actual Filtering and Ranking of query is done to pass it as input to SNMB. As the service brokers in Service Oriented Architecture, cloud brokers also rank the cloud resources that are published. Services can be preauthorized and prioritized in several groups such as price, availability, reliability, and security, etc. Ranking can be achieved all the way through user selection or historical service governance records. |
| Step 5. | UDDI Registry saves the Service Provider's details, details of Service Types and other parameters. Thus, the output from SNMB is compared and extracted from UDDI Registry. Service Providers are already registered through the facility of SPB and legal certificate is generated for future reference. |
| Step 6. | Finalized, more appropriate Service is forwarded to Consumer. |

## 5. Conclusion

This paper contributes a QoS framework for effective WS-Cloud Computing. The goal of the QoS in agent is to support advanced Web Service Discovery with QoS applied in registration, verification, certification, and confirmation. The uniqueness and consequence of WS-Cloud computing work are mentioned in two ways. Firstly, by introducing the idea of applying service matching, service ranks and service filtering, using QoS nonfunctional attribute that forms an effective cloud manageability property that gives rise to meet Consumer needs and while also organizing Cloud resources. Secondly, on perspective of Cloud Computing, contribution is the effective approach for facilitating QoS based Cloud Service Discovery, Service Negotiation, and Service Composition.